\documentclass[reprint,prb,superscriptaddress]{revtex4-1}

\usepackage{upgreek}
\usepackage[latin1]{inputenc}
\usepackage{subsupscripts}
\usepackage{subscript}

\usepackage{graphicx}
\usepackage{dcolumn}
\usepackage{bm}
\usepackage{color}

\newcommand{\mrm}{\mathrm}

\usepackage{booktabs}
\usepackage{amssymb}

\begin{document}

\preprint{}

\title{All-optical flow control of a polariton condensate using non-resonant excitation}
\author{Johannes Schmutzler}
\affiliation{Experimentelle Physik 2, Technische Universit\"at Dortmund, \mbox{D-44221 Dortmund, Germany}}

\author{Przemyslaw Lewandowski}
\affiliation{Department of Physics and CeOPP, University of Paderborn, Warburger Str. 100, D-33098 Paderborn, Germany}
\author{Marc A{\ss}mann}
\affiliation{Experimentelle Physik 2, Technische Universit\"at Dortmund, \mbox{D-44221 Dortmund, Germany}}
\author{Dominik Niemietz}
\affiliation{Experimentelle Physik 2, Technische Universit\"at Dortmund, \mbox{D-44221 Dortmund, Germany}}
\author{Stefan Schumacher}
\affiliation{Department of Physics and CeOPP, University of Paderborn, Warburger Str. 100, D-33098 Paderborn, Germany}
\author{Martin Kamp}
\affiliation{Technische Physik, Physikalisches Institut, Wilhelm Conrad R\"ontgen Research Center for Complex Material Systems,
Universit\"at W\"urzburg, D-97074 W\"urzburg, Germany}
\author{Christian Schneider}
\affiliation{Technische Physik, Physikalisches Institut, Wilhelm Conrad R\"ontgen Research Center for Complex Material Systems,
Universit\"at W\"urzburg, D-97074 W\"urzburg, Germany}
\author{Sven Höfling}
\affiliation{Technische Physik, Physikalisches Institut, Wilhelm Conrad R\"ontgen Research Center for Complex Material Systems,
Universit\"at W\"urzburg, D-97074 W\"urzburg, Germany}
\affiliation{SUPA, School of Physics and Astronomy, University of St Andrews, St Andrews, KY16 9SS, United Kingdom}

\author{Manfred Bayer}
\affiliation{Experimentelle Physik 2, Technische Universit\"at Dortmund, \mbox{D-44221 Dortmund, Germany}}
\affiliation{A. F. Ioffe Physical-Technical Institute, Russian Academy of Sciences, St Petersburg 194021, Russia} 

\date{15 April 2015}

\begin{abstract} 
The precise adjustment of the polariton condensate flow under incoherent excitation conditions is an indispensable prerequisite for polariton-based logic gate operations.
In this report, an all-optical approach for steering the motion of a polariton condensate using only non-resonant excitation is demonstrated. We create arbitrarily shaped functional potentials by means of a spatial light modulator, which allow for tailoring the condensate state and guiding a propagating condensate along reconfigurable pathways. Additional numerical simulations confirm the experimental observations and elucidate the interaction effects between background carriers and polariton condensates. 
\end{abstract}

\pacs{71.36.+c, 42.55.Px, 42.55.Sa, 73.22.Lp}


\maketitle

\section{Introduction}
Currently there is a lively debate concerning the role of all-optical logic circuits for future computation devices  \cite{Caulfield2010,Miller2010a,Miller2010b,Tucker2010, Snoke2013}.
In its course, major challenges for all-optical circuits to compete with state-of the art complementary metal-oxide-semiconductor (CMOS) technology regarding device footprint, energy consumption and production costs have been outlined \cite{Miller2010a,Miller2010b,Tucker2010}.
Altogether, all-optical circuits might be promising alternatives regarding heat dissipation and operation speed \cite{Caulfield2010,Snoke2013} and most importantly, they allow in principle for the dynamic design of an optical circuit by alteration of the applied electromagnetic fields \cite{Assmann2012}.

An appealing system for the realization of all-optical logic circuits are exciton-polaritons  in  semiconductor microcavities, which can occupy a single state in a macroscopic number and reveal several features of Bose-Einstein condensates (BECs) \cite{Kasprzak2006}. Moreover, dissipationless coherent propagation of polariton condensates over hundreds of microns \cite{Wertz2010,Nelsen2013}, frictionless flow \cite{Amo2009b}, a propagation speed on the order of $1~\%$ of the speed of light \cite{Wertz2012} and dispersionless propagation of polariton condensates \cite{Egorov2010a,Sich2012} highlight promising features of polaritons concerning logic gate operations.
Recently, several groups succeeded in the demonstration of a proof of principle transistor operation of polaritons \cite{Gao2012,Nguyen2013,Ballarini2013}. In Ref.~\onlinecite{Ballarini2013} even more sophisticated features such as cascadability of two transistors and logic gate operation were demonstrated.
However, the approaches presented in Refs.~\onlinecite{Gao2012,Nguyen2013} require lithographic patterning and are therefore strictly speaking not all-optical circuits. The resonant excitation scheme used in Ref.~\onlinecite{Ballarini2013} demands a careful choice of energy and angle of the excitation laser and beyond that, the number of laser beams impinging on the sample scales with the number of transistors cascaded. This might be a serious drawback for a large scale application. 

As an alternative one might consider non-resonant laser excitation for the realization of logic gates based on microcavity polaritons. Thereby, one has a large degree of freedom regarding the choice of excitation angle and energy. On the other hand, in contrast to resonant laser excitation, for an all-optical approach one needs to control the flow direction by means of optically created potentials. These potentials can be realized under non-resonant pumping by the simultaneous creation of background carriers, from which the condensate is repelled. While these potentials were exploited to a large extent for the realization of trapping geometries \cite{Tosi2012,Cristofolini2013,Askitopoulos2013,Dall2014}, only a discretization of the momentum distribution was shown so far for the case of nontrapping geometries \cite{Assmann2012}. 

In this report, we demonstrate a directed condensate flow over large distances on the order of $20~\upmu\mbox{m}$ using optically generated potentials. Furthermore, by a reconfiguration of the optical potential, we are able to recapture the condensate flow as well as to steer the condensate propagation in arbitrary directions. Moreover, our experimental results are confirmed by numerical simulations using a generalized Gross-Pitaevskii equation (GPE). Control of the condensate flow is an important milestone on the way towards a functional circuit architecture based on microcavity polaritons \cite{Keyes1985,Liew2008,Ortega2013,Sigurdsson2014}.

\section{Experimental details}
We investigate a planar GaAs-based microcavity with a quality factor of about $20000$ and a Rabi splitting of $9.5~\mbox{meV}$. The structure of the sample is as follows: Four GaAs quantum wells are placed in the central antinode of the electric field confined by two distributed Bragg reflector (DBR) structures in a $\uplambda$/2-cavity. The upper (lower) DBR structure consists of 32 (36) alternating layers of Al\textsubscript{0.2}Ga\textsubscript{0.8}As and AlAs.
The sample is mounted in a helium-flow cryostat, measurements are
performed at $10~\mbox{K}$. For non-resonant optical excitation a
femtosecond-pulsed Titanium-Sapphire laser (repetition rate
$75.39~\mbox{MHz}$) with central wavelength at $727~\mbox{nm}$
($1705~\mbox{meV}$) is used. For generation of the optical potentials, the laser beam is divided using a beamsplitter. The shape of the first beam is Gaussian with a full width at half maximum (FWHM) of $2~\upmu\mbox{m}$ on the sample, the shape of the second laser beam can be arbitrarily modulated using a spatial light modulator (SLM). Both beams are focused under normal incidence onto the sample using a microscope objective (numerical aperture $0.42$), the time delay between the beams is less than $2~\mbox{ps}$.
For detection a liquid nitrogen-cooled CCD-camera behind a monochromator is used. For two-dimensional imaging in real space and Fourier-space, respectively, the monochromator is operated in zeroth order. In that case, the spectral resolution is provided by a bandpassfilter with a FWHM of $1~\mbox{nm}$.
All experiments are performed at an exciton-cavity detuning of $-21.7~\mbox{meV}$, which corresponds to a photonic fraction of $96~\%$ of the lower polariton (LP) at zero in-plane wavevector.

\section{Theoretical model}
\label{Theory}

In order to obtain a better understanding of the interaction between the optically created potentials and the polariton condensates we performed numerical simulations. Our analysis is based on a mean-field description of the coherent polariton field coupled to two incoherent background carrier/exciton reservoirs, which represent an active and an inactive reservoir, respectively  \cite{Veit2012,Lagoudakis2011}. 
The implementation of an inactive reservoir reflects the relaxation process from background carriers induced by the non-resonant pumping process  towards an active exciton reservoir, which feeds the coherent polariton field through stimulated scattering. 
Our theoretical description captures the coupled spatio-temporal dynamics of (a) the coherent polariton field/condensate $\Psi$, (b) the active exciton reservoir $n_A$ feeding the condensate  and (c) the inactive reservoir $n_I$ created in the optical pumping process and feeding the active reservoir. The equations of motion read as follows:
\begin{eqnarray}
i \hbar \dot{\Psi} & = & \left( \mathbb{H} - \mathrm{i} (\gamma_p - {{\gamma}\over{2}}n_A) + V_{d} \right) \Psi \\ 
& + & \left( \alpha_1   \vert \Psi \vert^2    +  \alpha_2 n_A + \alpha_3 n_I \right) \Psi - i \Lambda (n_A+n_I)\mathbb{H} \Psi \nonumber \\
\dot{n}_A & = & {{1}\over{\hbar}} \left(\tau n_I - \gamma_A n_A  -\gamma \vert \Psi \vert^2 n_A \right)  \\
\dot{n}_I & = & {{1}\over{\hbar}} \left( -\tau n_I - \gamma_I n_I\right).
\end{eqnarray} 

\begin{figure}[!htb]
\centering
\includegraphics[width=0.95\linewidth]{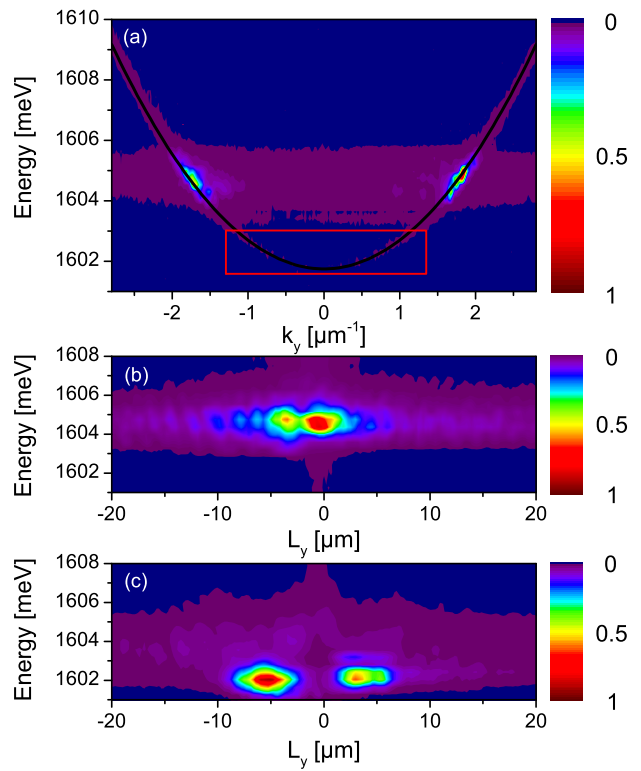}
\caption{(Color online) (a) Fourier-space image for an excitation power level $P=8~P_{\mrm{thr}}$ under excitation with the Gaussian laser spot only. The red rectangle indicates the selected spectral region for the experiments operating the monochromator in zeroth order. (b) Corresponding real-space image for $P=8~P_{\mrm{thr}}$. (c) Real-space image of the condensate emission for the  excitation profile presented in Fig.~\ref{Fig2}~(a). The excitation power levels for the Gaussian spot is $P=18~P_{\mrm{thr}}$ and for the circle shaped pattern $P=6.5~P_{\mrm{thr}}$, respectively. Threshold power levels are determined by exciting either with the Gaussian spot or the circle shaped pattern only.}
\label{Fig1}
\end{figure}

$\Psi$, $n_A$, and $n_I$ are defined in the two-dimensional $x,y$-plane. Eq.~(1) describes the dynamics of the coherent polariton field and is of the type of a modified Gross-Pitaevskii equation. The coherent part can be derived from a microscopic semiconductor theory in mean-field approximation and parabolic approximation for the lower polariton branch \cite{Luk2013}; spin degrees of freedom\cite{Ardizzone2013} are not considered here. Eqs.~(2-3) describe the dynamics of the active and inactive reservoir, respectively. 
In Eq.~(1), $\mathbb{H} = -{{\hbar^2}\over{2m_p}} \Delta$ accounts for the free propagation of polaritons with effective mass $m_p = 0.35 \cdot 10^{-4}~m_e$. The polariton field $\Psi$ is replenished by the active reservoir with $\gamma=0.004~\mbox{meV}\upmu\mbox{m}^2$. A disorder potential $V_d$ is included with a spatial correlation length of 1 $\upmu$m and a root mean square (rms) amplitude of $0.2~\mbox{meV}$.
A repulsive Coulomb interaction is given by $\alpha_1= 0.0024~\mathrm{meV\upmu m^2}$ (Refs.~\onlinecite{Manni2011b, Anton2013, Lagoudakis2011}) for interactions between polaritons and by
 $\alpha_2=\alpha_3= 0.008\,\mathrm{meV\upmu m^2}$ for the polariton-reservoir interaction. The last term in Eq.~(1) with $\Lambda = 0.00025~\upmu\mbox{m}^{-2}$ mimics a relaxation term as it drives the polariton system to on average lower kinetic energies in spatial regions where interaction with the reservoir densities $n_A$ and $n_I$ occurs \cite{Wertz2012som}. This approach is based on Ref.~\onlinecite{Choi1998} and related to an early work of L. P. Pitaevskii \cite{Pitaevskii1959} to describe damping in superfluid helium. The physical origin of energy dissipation of polaritons is scattering with a thermalized reservoir of excitons that occupy mainly low-energy states. When polariton-exciton scattering occurs promoting excitons to higher energy states, energy conservation leads to loss of energy of condensed polaritons. Then, the excess energy attained by the excitons can quickly be dissipated through phonon emission. In Eqs.~(2-3), the active reservoir is fed by the inactive reservoir with $\tau=0.1$~meV.
Radiative losses are $\gamma_p=0.1$ meV,  $\gamma_A=0.01~\mathrm{meV}$ (Ref.~\onlinecite{Haug2014}) and $\gamma_I=0.0013\,\mathrm{meV}$ (Ref.~\onlinecite{Lagoudakis2011}) for $\Psi$, $n_A$ and $n_I$, respectively. 

In the simulations, Eqs.~(1-3) are solved explicitly in time on a two-dimensional grid in real space. Initially, the active reservoir density $n_A$ is set to zero. The optical excitation is assumed instantaneously in the calculations (experimentally, excitation is far above the gap on a fast $100\,\mathrm{fs}$ timescale) and is included through the spatial profile of inactive reservoir density $n_I$ used as initial condition. The maximum initial reservoir density is $\mathrm{max}(n_I)=1.1 \cdot\,\mathrm{10^{11}~cm^{-2}}$. A small random field with random phase and amplitude is assumed for the coherent polariton field $\Psi$ to trigger the stimulated feeding after optical excitation. This initial random field has a spatial correlation length of $1~\upmu\mbox{m}$ and a rms amplitude of $10^{7}$ cm$^{-2}$. If only a single run is studied in the calculations, interference phenomena can be visible (e.g., for the polariton flow leaving the half-open trap as discussed in Sec.~\ref{Results and discussion}). However, these interference effects are not observed in the experiment where the  photoluminescence is averaged over multiple emission events for the pulsed excitation. Taking this into account, we average the condensate distribution over multiple calculations (30 leads to good convergence) for a fixed disorder configuration but with different initial random fields for $\Psi$.

\section{Results and discussion}
\label{Results and discussion}

Fig.~\ref{Fig1}~(a) shows a typical dispersion for an excitation power level $P$ several times higher than the threshold power $P_{\mrm{thr}}$ for polariton condensation, when only the Gaussian laser spot is used. Clearly, the main emission at wavevectors of $|k_y|\approx1.8~\upmu\mbox{m}^{-1}$ occurs blueshifted by $3~\mbox{meV}$ with respect to the lowest energy state of the LP at zero in-plane wavevector. The observed blueshift arises from repulsive interactions between the condensed polaritons and background carriers created by non-resonant pumping and from polariton-polariton interaction. These mechanisms generate an antitrapping potential and cause a ballistic acceleration of the condensed polaritons \cite{Wouters2008}.
This finding is further elucidated by the corresponding real space spectrum, where a pronounced  ballistic propagation over a distance of several $10~\upmu\mbox{m}$ can be seen [Fig.~\ref{Fig1}~(b)]. However, as a two-dimensional polariton system is investigated here, this propagation occurs omnidirectional, radially symmetric with respect to the Gaussian excitation laser spot. 

\begin{figure}[!htb]
\centering
\includegraphics[width=\linewidth]{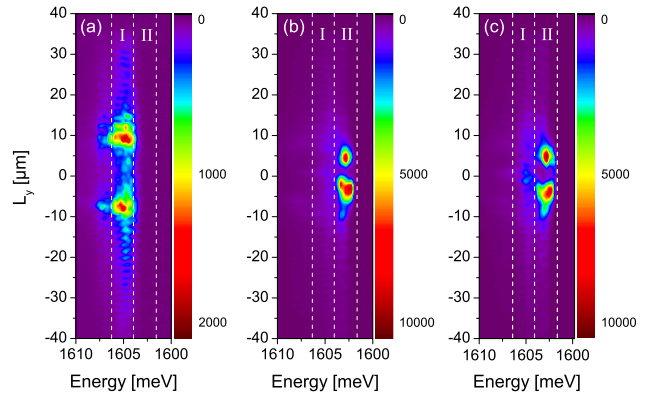}
\caption{(Color online) Energy resolved cross-sections of the polariton condensate using a ring shaped laser spot with excitation power $P=2.3~P_{\mrm{thr}}$ and a Gaussian laser spot with different excitation power levels of   $P=0$ (a),  $P=0.75~P_{\mrm{thr}}$ (b) and $P=4.25~P_{\mrm{thr}}$ (c). For quantifying the trapping efficiency, the overall count number within region I and II was integrated.}
\label{Additional_figure1}
\end{figure}

\begin{figure}[!htb]
\centering
\includegraphics[width=\linewidth]{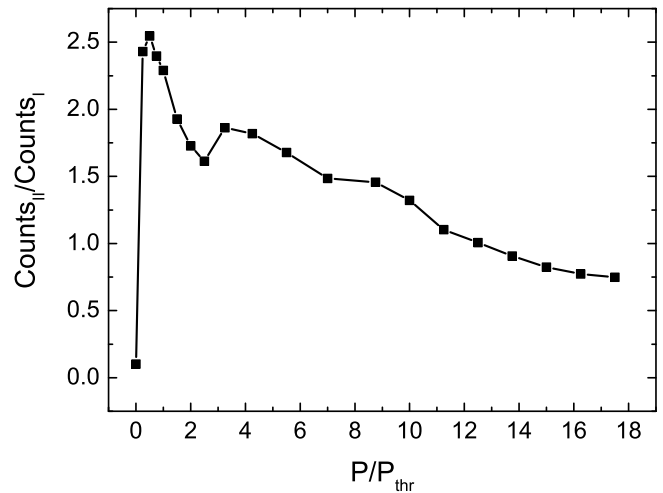}
\caption{Trapping efficiency for different excitation power levels of the Gaussian laser spot. The trapping efficiency is determined by dividing the count number of region II by the count number of region I as shown in Fig.~\ref{Additional_figure1}.}
\label{Additional_figure2}
\end{figure}
 
\begin{figure}[!htb]
\centering
\includegraphics[width=0.97\linewidth]{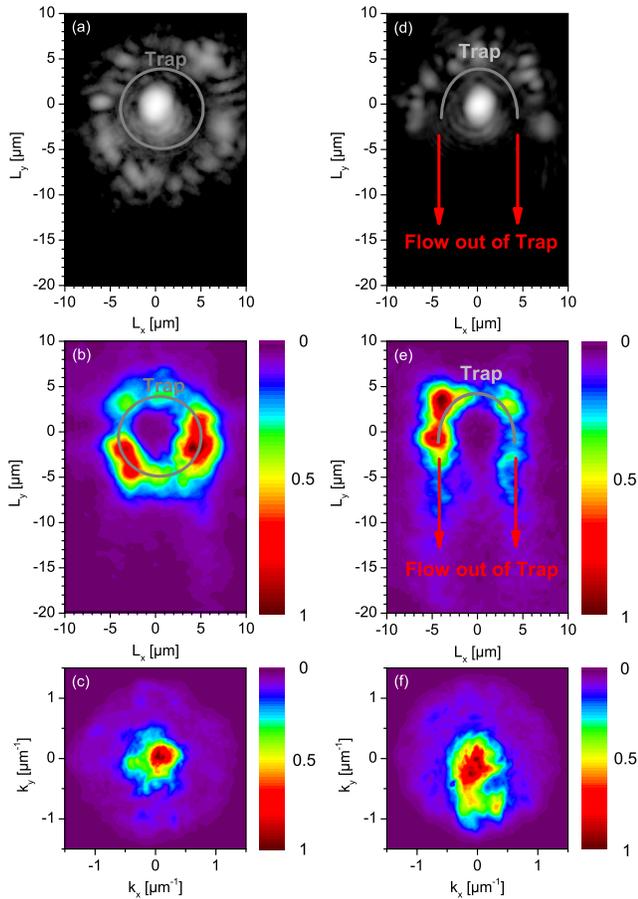}
\caption{(Color online) (a) Laser excitation pattern for the trapping geometry. The excitation power levels for the Gaussian spot is $P=18~P_{\mrm{thr}}$ and for the circle shaped pattern $P=6.5~P_{\mrm{thr}}$. (b) Twodimensional real-space image of the trapped condensate. (c) Corresponding Fourier-space image of the trapped condensate. (d) Laser excitation pattern for the source of a directed condensate flow. The excitation power levels for the Gaussian spot is $P=18~P_{\mrm{thr}}$ and for the semicircle shaped pattern $P=11.3~P_{\mrm{thr}}$. (e) Twodimensional real-space image of the trapped condensate and the directed condensate flow escaping from the trap. (f) Corresponding Fourier-space image for the situation of panel (e).}
\label{Fig2}
\end{figure}

\begin{figure}[!htb]
\centering
\includegraphics[width=0.97\linewidth]{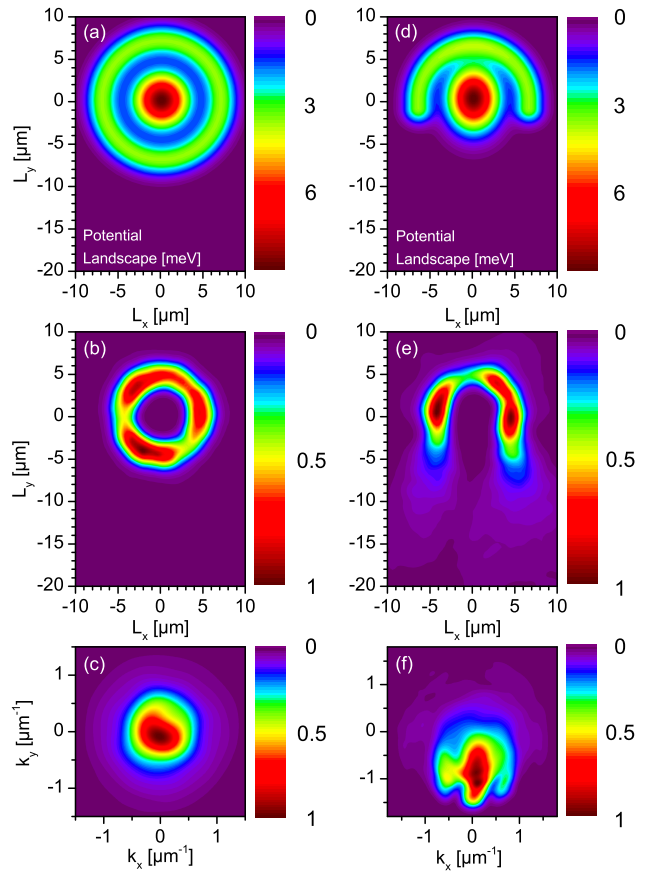}
\caption{(Color online) Applied potential landscape for the simulation of the trapping geometry (a) and of the source of the directed condensate flow (d). Corresponding time-integrated  calculated condensate distribution in real-space [(b) and (e)] and in k-space [(c) and (f)].}
\label{Fig3}
\end{figure}

\begin{figure}[!htb]
\centering
\includegraphics[width=0.96\linewidth]{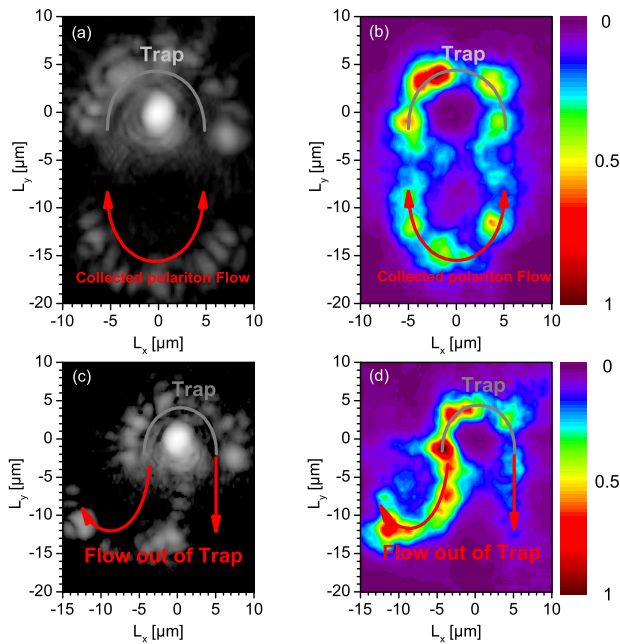}
\caption{(Color online) (a) Laser excitation pattern for the collector geometry. The excitation power levels for the Gaussian spot is $P=18~P_{\mrm{thr}}$ and for the SLM generated pattern $P=5.9~P_{\mrm{thr}}$. (b) Twodimensional real-space image of the recollected condensate. (c)  Laser excitation pattern for bending the condensate flow. The excitation power levels for the Gaussian spot is $P=18~P_{\mrm{thr}}$ and for the SLM generated pattern $P=6.2~P_{\mrm{thr}}$. (d) Twodimensional real-space image of the curve-shaped condensate.}
\label{Fig4}
\end{figure}

\begin{figure}[!t]
\centering
\includegraphics[width=0.96\linewidth]{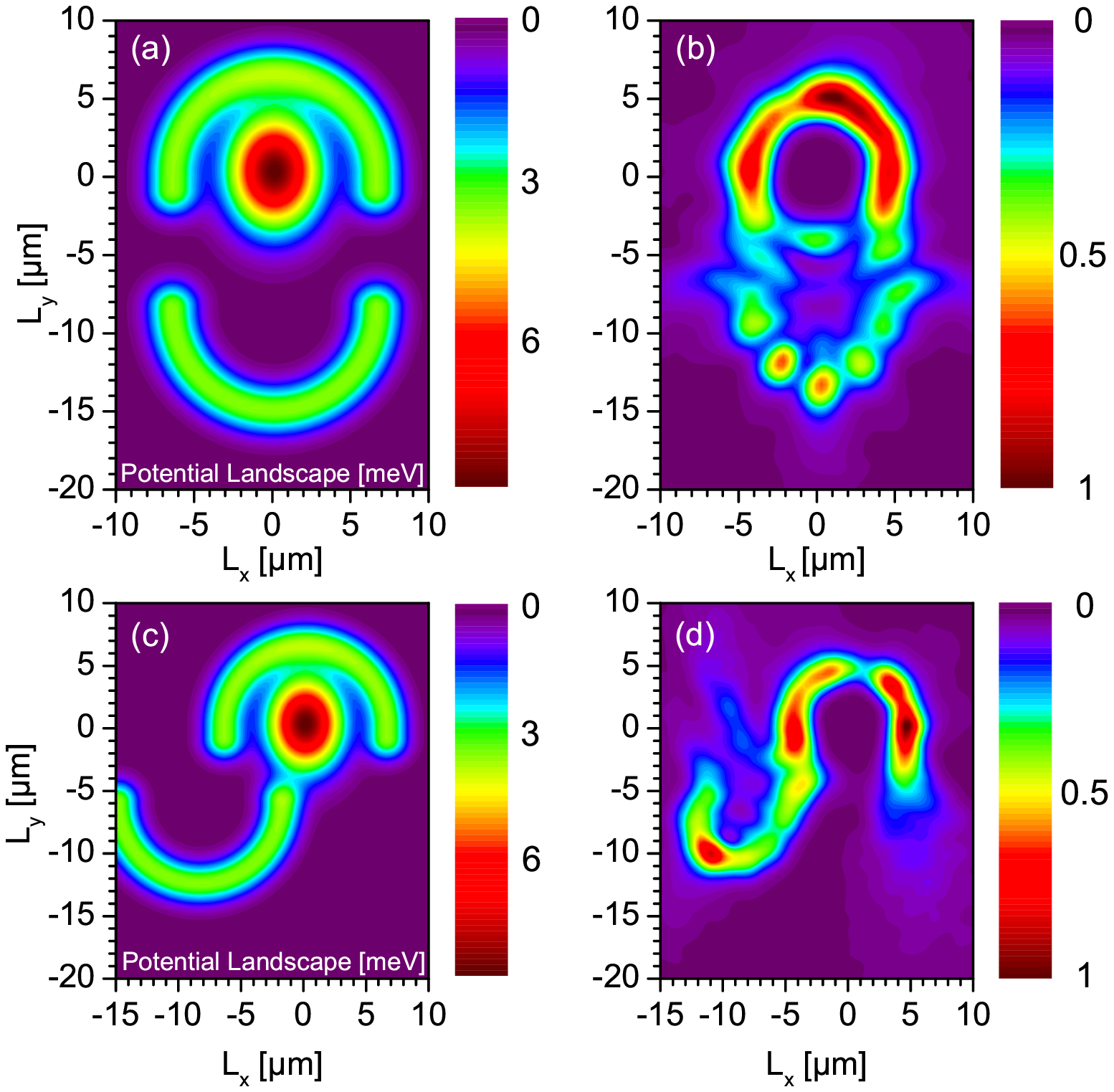}
\caption{(Color online) 
Applied potential landscape for the simulation of the collector geometry (a) and of the bending of the condensate flow (c). Corresponding time-integrated  calculated condensate distribution in real-space [(b) and (d)].}
\label{Fig5}
\end{figure} 

As a first step towards controlling the spread of the condensate, we trap the condensate  by applying a ring shaped laser spot, ca. $10~\upmu\mbox{m}$ in diameter, centrosymmetric around the Gaussian spot using the second laser beam modulated by the SLM. A typical excitation profile for this situation is depicted in Fig.~\ref{Fig2}~(a). The condensates excited by this laser pattern are trapped due to the repulsive interaction with background carriers generated at the position of the laser pattern. Fig.~\ref{Fig1}~(c) shows a spectrally resolved cross-section in $L_y$ direction, where $L_y=0~\upmu\mbox{m}$ indicates the position of the Gaussian laser spot. In this case the main emission is centered at $1602~\mbox{meV}$, $3~\mbox{meV}$ lower in energy compared to the case of excitation with the Gaussian laser spot only.
 
At this point it should be mentioned that the overall trapping efficiency is highly dependent on the sample position due to disorder effects as well as on the chosen excitation power levels for both laser beams. 
We have performed additional measurements with varying excitation power levels of the Gaussian laser spot at a slightly  different sample position with the same detuning. Here, we have measured energy-resolved cross-sections of the polariton condensate in real space. Fig.~\ref{Additional_figure1} shows exemplarily three cross-sections for different excitation powers of the Gaussian laser spot at a fixed excitation power $P=2.3~P_{\mrm{thr}}$ of the ring shaped laser spot.  As expected, no trapping is observed without additional excitation using the Gaussian laser spot in the center of the circular shaped spot [Fig.~\ref{Additional_figure1}~(a)].

A pronounced trapping of the condensate is already evident for excitation power levels below threshold of the Gaussian laser spot [Fig.~\ref{Additional_figure1}~(b)], which is maintained also for higher excitation power [Fig.~\ref{Additional_figure1}~(c)]. For a comparison of the trapping efficiency at different excitation power of the Gaussian laser spot, the overall count rate within the high energy region of the free propagating condensate (region I, indicated by dashed lines in Fig.~\ref{Additional_figure1}) and the low energy region of the trapped condensate (region II, indicated by dashed lines in Fig.~\ref{Additional_figure1}) was determined. We regard the ratio of the count number of region II and region I as a reliable measure to quantify the trapping efficiency, which is shown in Fig.~\ref{Additional_figure2}.
Clearly, a pronounced decrease of the trapping efficiency can be observed when the excitation power of the central Gaussian laser spot is much larger compared to the ring shaped laser spot. In this situation, the condensate which is ballistically accelerated away from the center cannot be trapped by the potential landscape induced by the ring shaped laser spot. Consequently,  the excitation power levels of both laser spots should be chosen in the same order of magnitude with respect to condensation threshold to allow for an efficient energy relaxation and prevent escape from the trap.

For the further experiments we select a spectral region of ca. $1~\mbox{meV}$ width using an interference filter, covering wavevectors $|k_{||}|<1.3~\upmu\mbox{m}^{-1}$ only, which is indicated by the red rectangle in Fig.~\ref{Fig1}~(a). 
In Fig.~\ref{Fig2}~(b) the trapped condensate can be seen, which appears in a donut shape due to the centrosymmetric shape of the applied optical potential [Fig.~\ref{Fig2}~(a)]. Furthermore, the wavevector distribution is centered around zero momentum, as expected for a trapped condensate [Fig.~\ref{Fig2}~(c)].

For the numerical simulation of the trapping geometry we use a centrosymmetric shaped potential landscape as depicted in Fig.~\ref{Fig3} (a), which results in a trapping of the condensate with $\mathbf{k}\approx 0$ [Fig.~\ref{Fig3}~(b) and (c)] due to interaction with two reservoirs of active and inactive background carriers, in accordance with the experimental observations [Fig.~\ref{Fig2}~(b) and (c)]. 

As outlined in Sec.~\ref{Theory}, the inclusion of an active and an inactive reservoir in our theoretical description, phenomenologically captures the sequential built-up of a coherent polariton field for pulsed excitation far above the band gap. Only a small amount of excitons fulfills the phase-matching condition for stimulated scattering into the condensate and can therefore contribute to the build-up of the condensate directly. There is, however, also a large amount of background excitations, which cannot directly scatter into the condensate and decays much slower compared to the polariton condensate. The latter was recently demonstrated by terahertz spectroscopy \cite{Menard2014}. From this observation it becomes clear, that a differentiation between active background excitons, which can directly populate the condensate by stimulated scattering, potentially on a very short timescale, and an inactive reservoir of excitations, which decays on a longer timescale, is required for an accurate description of the two mechanisms gain and trapping, provided by the background excitations. For the pulsed excitation studied here we found it indeed necessary to model the active and inactive reservoir separately in the numerical simulations. Only then we could  obtain at the same time condensate formation but also trapping of the condensate through the reservoir generated potential as observed in the experiments. In a single-reservoir model, once the condensate builds up the reservoir is completely consumed through the stimulated scattering such that no trapping potential remains. Consequently, two background carrier reservoirs are required  for an accurate description of the condensate dynamic under pulsed excitation conditions, which has also been reported by other groups \cite{Lagoudakis2010, Lagoudakis2011,DeGiorgi2014}. We note that the situation may be different when continuous wave pump excitation is studied such that a quasi-steady state is reached and with a reservoir of background excitations that is persistently fed by the cw pump source \cite{Manni2011b,Lagoudakis2008}.
 
To exploit the trapping geometry for the generation of a directed condensate flow, we have opened the trap by applying a semicircle shaped potential instead of a full circle [Fig.~\ref{Fig2}~(d)], the corresponding potential landscape used for our simulations is presented in Fig.~\ref{Fig3}~(d). After generation, the polariton condensate can leave the trap by flowing through the aperture at the bottom in $L_y$ direction [Fig.~\ref{Fig2}~(e)]. Due to the repulsive interaction with the reservoir excitons, polaritons are accelerated away from the excitation point such that a directed polariton flow from the trap is clearly visible in the experiment [Fig.~\ref{Fig2}~(e)], whereas it is less pronounced in the numerical simulations [Fig.~\ref{Fig3}~(e)]. We note that the quantitative appearance of the computed photoluminescence depends on the specific disorder configuration. As the disorder potential can not be extracted from the experiment, only qualitative agreement between measured and calculated photoluminescence is obtained here.
The pronounced directed propagation is also evident in Fourier-space, where the wavevector distribution is significantly relocated towards negative values of $k_y$ [Fig.~\ref{Fig2}~(f) and Fig.~\ref{Fig3}~(f)] in contrast to the situation of the closed trap [Fig.~\ref{Fig2}~(c) and Fig.~\ref{Fig3} (c)]. Therefore this geometry acts as a source for a directed condensate flow.

Once a directed flow is realized, one can further steer and manipulate this directed condensate arbitrarily by modifying the potential landscape. In the following we demonstrate exemplarily two scenarios.  By applying an additional semicircle shaped laser pattern in roughly $10~\upmu\mbox{m}$ distance from the source [Fig.~\ref{Fig4}~(a)], the condensate flow is recollected and transferred to an oval shaped standing wave pattern [Fig.~\ref{Fig4}~(b)] as a consequence of reflection, amplification and interference, which is reproduced well by our simulations [Fig.~\ref{Fig5}~(a) and (b)]. 
Furthermore, if an opposing semicircle  shaped potential with respect to the source is applied [Fig.~\ref{Fig4}~(c)], one can generate a curve shaped condensate flow [Fig.~\ref{Fig4}~(d)], which is consistent with our calculations [Fig.~\ref{Fig5}~(c) and  Fig.~\ref{Fig5}~(d)]. 

In both scenarios the optically created potential interacts twofold with the incoming condensate: Firstly the incoming flow is redirected due to repulsive Coulomb interaction with the background carriers, and secondly the potential barrier operates also as a gain medium, which gives rise to strong condensate emission in up to $15~\upmu\mbox{m}$ distance from the source of the directed condensate flow. Nevertheless, the barrier remains separated in space from the condensate flow, which might be beneficial concerning the loss of coherence of a condensate mediated by the local presence of background carriers \cite{Schmutzler2014}.

\section{Conclusion and outlook}
In conclusion, we have demonstrated control of the polariton flow using reconfigurable optically induced potentials. Firstly, we have presented a feasible approach for the generation of directed condensate propagation. Furthermore, a manipulation of this flow, e.g. a recollection of the condensate and a constraint on a curve shaped trajectory, has been demonstrated. In addition we have reproduced our experimental results in terms of a generalized GPE. The control of the condensate flow is a prerequisite for further, more sophisticated investigations, e.g. scattering experiments of polariton condensates, and  might pave the way for the realization of optically generated printed circuit boards for polariton-based logic circuits.

\section*{Acknowledgements}
The Dortmund group acknowledges support from the Deutsche Forschungsgemeinschaft (Grants 1549/18-1 and SFB TRR 142). 
The Paderborn group is grateful for financial support from the Deutsche Forschungsgemeinschaft (GRK 1464, SFB TRR 142 and SCHU 1980/5-1) and acknowledges a grant for computing time from the PC$^2$ Paderborn Center for Parallel Computing.  The group at W\"urzburg
University acknowledges support from the State of Bavaria.

%
\end{document}